\documentclass[aps,prl,showpacs]{revtex4}

\newcommand{\be}{\begin{equation}}
\newcommand{\ee}{\end{equation}}

\begin{document}

\large

\title{Arbitrary Spin Galilean Oscillator}

\author{C. R. Hagen\footnote{mail to hagen@pas.rochester.edu} }
\affiliation{Department of Physics and Astronomy\\
University of Rochester\\
Rochester, N.Y. 14627-0171}

\begin{abstract}
The so-called Dirac oscillator was proposed as a modification of the free Dirac equation which reproduces many of the properties of the simple harmonic oscillator but accompanied by a strong spin-orbit coupling term.  It has yet to be extended successfully to the arbitrary spin $S$ case primarily because of the unwieldiness of general spin Lorentz invariant wave equations.  It is shown here using the formalism of totally symmetric multispinors that the Dirac oscillator can, however, be made to accommodate spin by incorporating it into the framework of Galilean relativity.  This is done explicitly for spin zero and spin one as special cases of the arbitrary spin result.  For the general case it is shown that the coefficient of the spin-orbit term  has a $1/S$ behavior by techniques which are virtually identical to those employed in the derivation of the $g$-factor carried out over four decades ago.
\end{abstract}



\maketitle


\section{I Introduction}

In 1989 Moshinski and Szczepaniak \cite{Moshinsky} proposed a formulation of the simple harmonic oscillator (SHO) which was based on the Dirac equation.  They argued that since the first order differential Dirac equation was linear in the momentum $p$ but led to a quadratic dependence on $p$ in the relativistic energy-momentum relation, it was natural to expect that the quadratic dependence  on the spatial coordinate in the SHO could emerge by a linear dependence on it in the corresponding Dirac equation.  Thus they proposed that the momentum four-vector $p^\mu  ( \mu=0, 1, 2, 3)$ be replaced by $({\bf p}-iM\omega{\bf r}\beta, p^0)$   where $M$ is the particle mass, $\omega$ a frequency parameter, and $\beta$ the zero component of the Dirac set $\gamma^\mu$.  This leads one to consider the modified Dirac equation
$$E\psi=[\beta {\bf \gamma} \cdot ({\bf p}-iM\omega {\bf r}\beta)+M\beta]\psi$$
for the four component wave function $\psi$ in units in which $\hbar=c=1$.  Upon writing $\psi$ in terms of two two-component wave functions $\psi_1$ and $\psi_2$ it is readily shown that $\psi_1$ satisfies a Schr\"{o}dinger equation with eigenvalues given by
$${E^2-M^2\over 2M}=\omega[2n+\ell-{\bf L}\cdot{\sigma}]$$
where $n=0,1,2,...$, $\ell$ is the angular momentum quantum number, ${\bf L}$ is the angular momentum operator, and ${\bf \sigma}$ are the usual Pauli spin matrices.  In the low energy limit the left hand side of this relation becomes the  nonrelativistic energy ${\cal E}$ and one has
$${\cal E}= \omega[ 2n +(\ell +1)^2-(j+{1\over 2})^2]$$
where $j=\ell\pm {1\over 2}$ is the usual total angular momentum quantum number. .

This is, of course, quite similar to the SHO result.  The sole differences are the absence of the zero point energy ${3\over 2}\omega$ and the presence of a strong spin-orbit coupling term.  Since the Dirac oscillator has been the object of considerable study following its first inception, it is noteworthy that there has not been a successful extension of it to include other spin values.  This is no doubt attributable at least in part to the considerable complications which ensue when the full formalism of arbitrary spin Lorentz covariant theories is invoked.  Since, however, the intereresting properties of the Dirac oscillator are obtained already in the nonrelativistic limit, it would seem that there is much to be learned from an application of the much simpler formalism which describes spin in the Galilean relativistic limit. 

In this work the extension of the Dirac oscillator to arbirary spin is carried out invoking only the principles of Galilean relativity.  It is well to remind a possibly skeptical  reader that Galilean relativity can be a very powerful tool and was in fact responsible for the proof of a longstanding conjecture that the $g$-factor of a spin $S$ particle is $1/S$ \cite{Hagen}.   Since, however, there is considerably less familiarity with the Galilean spin one-half theory than with the Dirac equation, the following section is devoted to a rederivation of the principal results of the Dirac oscillator within the Galilean framework.  Having accomplished this introduction, section {\bf III} extends the newly derived Galilean spin one-half oscillator to the spinless and spin one cases using the ideas of symmetrized (and antisymmetrized) multispinors.  The general spin $S$ formalism is then presented in section {\bf IV} and the result obtained that just as in the Dirac oscillator case the only change relative to the SHO is the absence of the zero point energy and the presence of a strong spin-orbit coupling term.  A secondary result reminiscent of the $g$-factor derivation is the emergence of a {1/S} dependence for the coefficient of that coupling term.  It is also shown that the latter result is explicitly dependent upon the assumption that one is dealing with a so-called minimal theory.  That is to say, the formalism is maximally parsimonious in the number of wave function components which it employs.

\section{II A Galilean spin one-half oscillator}

It was shown by L\'{e}vy-Leblond \cite{Levy-Leblond} in the spin one-half case that what was long taken to be a triumph of the Dirac theory in predicting the correct $g$ factor for the electron is in fact merely a consequence of the requirement that the wave equation be Galilean covariant and of first order in all derivatives.  One finds in particular that the assumed form 
$$G\psi=(Ai{\partial\over \partial t}+{\bf B}\cdot {1\over i}{\bf  \nabla}+C)\psi=0$$
is Galilean  covariant for the following form of the matrices
$$A={1\over 2}(1+\rho_3)$$
$$  B_i=\rho_1 \sigma_i$$
$$ C=M(1-\rho_3)$$
where the matrices $\rho_i$ and $\sigma_i$ are two commuting sets of Pauli matrices used to span the 4 $\times$ 4 dimensional spinor space.  The transformation law for $\psi$ corresponding to the Galilean transformation 
$${\bf x'}=R{\bf x}+{\bf v}t+{\bf a}$$
$$t'=t+b,$$
is
$$\psi'({\bf x'},t')=\exp[if({\bf x},t)]\Delta^{1/2}({\bf v},R)\psi({\bf x},t)$$
where 
$$f({\bf x}, t)={1\over 2}Mv^2  t+M{\bf v}\cdot R{\bf x} $$
and

$$ \Delta ^{1/2} ( {\bf v}, R) =  \left ( \begin{array}{cc}
D^{1/2}(R) &  0 \\
-{1\over 2}{\bf \sigma} \cdot {\bf  v}D^{1/2}(R)  & \;\;\; D^{1/2}(R)
\end{array} \right )$$
with  $D^{1/2}(R)$ being the usual two-dimensional representation of spin one-half which acts in the space of the $\sigma$ matrices.  The transformation law for $\psi$ displays the important fact that the upper components of $\psi$ do not mix with the lower components under a pure Galilean transformation.

Upon writing $\psi$ in terms of the two-component spinors $\phi$ and $\chi$
$$\psi= \left( \begin{array}{c}
\phi \\
\chi \end{array}\right)$$
the wave equation 
$$G\psi=0$$
yields
$$i{\partial \over \partial t}\phi+{\bf \sigma}\cdot{1\over i}{\nabla}\chi=0$$
$$\\{\bf\sigma}\cdot {1\over i}\cdot {\nabla}\phi+2M\chi=0.$$
In order to construct the analog of the Dirac oscillator one might be tempted to use the matrix $A={1\over 2}(1+\rho_3)$ as the appropriate Galilean equivalent of the Dirac matrix $\beta$.  Since, however, the part of $A$ which does not involve $\rho_3$ would merely constitute a gauge transformation on $\psi$, one disregards that term and replaces ${\bf p}$ by ${\bf p}-iM\omega {\bf r}\rho_3$.

Noting that the equation for $\phi$ involves a time derivative and is thus a true equation of motion, one can proceed to eliminate $\chi$ by the constraint equation
$$2M\chi= -{\bf \sigma}\cdot({\bf p}-iM\omega {\bf r})\phi$$
 to obtain a Galilean version of the spin one-half Dirac oscillator.   This yields the Schr\"{o}dinger-like equation for $\phi$
$${\cal E}\phi={1\over 2M}{\bf \sigma}\cdot({\bf p}+iM\omega {\bf r}){\bf \sigma}\cdot({\bf p}-iM\omega {\bf r})\phi$$
which readily reduces to 
$${\cal E}\phi=[{1\over 2M}{\bf p}^2+{1\over 2}M\omega^2{\bf r}^2-{3\over 2}\omega -\omega {\bf L}\cdot{\bf \sigma}]\phi.$$
Making use of the usual result for the energy levels of the SHO in spherical coordinates (namely, $\omega [2n+\ell+{3\over2}]$ where $n,\ell=0,1,2...)$, this yields the eigenvalue spectrum for the spin one-half Galilean oscillator as 
$$ {\cal E}_{nlj} = \omega  \left\{  \begin{array}{ll}
2n & \;\;\;  
      \mbox{$j=\ell+{1/2}$}\\
2n+2\ell+1 & \;\;\;  
 \mbox{$j=\ell -{1/2}$}.
\end{array}
\right. $$
This result is in exact agreement with that obtained in the nonrelativistic limit of the usual Dirac oscillator.  Specifically, it reproduces the two principal features of the latter--namely, the elimination of the zero point energy and the emergence of a strong spin-orbit coupling term.  It is therefore entirely appropriate to consider it to be the Galilean version of the Dirac oscillator.  The remainder of this work is devoted to the extension of this approach to the arbitrary spin case.   

\section{III Spin zero and spin one Galilean oscillators}

This paper seeks to extend the spin one-half Galilean result for the Dirac oscillator to arbitrary spin by using the techniques associated with symmetrized multispinors.  Thus one employs for the spin one case a second rank spinor $\psi_{ab}=\psi_{ba}$ which is to be described by the Lagrangian 
$${\cal L}={1\over 2}\psi^*_{ab}[G_{aa'}\Gamma_{bb'}+\Gamma_{aa'}G_{bb'}]\psi_{a'b'}.$$
The Galilean invariance of this Lagrangian follows from the properties of the matrix operator $G$ together with the fact that 
$$\Gamma \equiv  {1\over 2}(1+\rho_3)$$
is a matrix which is Galilean invariant.  This is an immediate consequence of the observation made earlier to the effect that upper components do not mix with lower ones under Galilean boosts.  Upon making the replacement of ${\bf p}$ by ${\bf p}-iM\omega {\bf r}\rho_3$ in $G$ there follows the spin one version of the Galilean oscillator.

The content of the implied equation
$$[G_{aa'}\Gamma_{bb'}+\Gamma_{aa'}G_{bb'}]\psi_{a'b'}=0$$
is made more transparent by displaying it in terms of its scalar and vector components. To this end one writes the symmetric bispinor $\psi_{ab}$ as 
$$\psi={1\over \sqrt 2}[{\bf X}\cdot {\bf \sigma}{1\over 2}(1+\rho_3)+{\bf Y}\cdot{\bf \sigma}\rho_1+Z\rho_2]\sigma_2.$$
One notes that although the symmetric bispinor $\psi$ should in principle have ten components (i.e., three more than implied by the seven objects ${\bf X}, {\bf Y}, Z$), one finds that the other three symmetric matrices given by ${1\over 2}(1-\rho_3)\sigma_i\sigma_2$ do not contribute to $\cal L$.  The Lagrangian ${\cal L}$ can now be written in terms of a trace as 
$${\cal L}={\rm Tr}G\psi\Gamma\psi^*.$$

Upon carrying out the trace one obtains 

\begin{eqnarray*}
{\cal L} & = &
{\bf X^*}\cdot i{\partial\over \partial t}{\bf X}+
{\bf X^*}\cdot {\bf \nabla}Z-Z^*{\bf \nabla}\cdot{\bf X}
+{\bf X^*}\cdot{\bf \nabla}\times {\bf Y}+{\bf Y^*}\cdot {\bf \nabla}\times {\bf X}  \\
&   & + 2M({\bf Y^*}\cdot {\bf Y}+Z^*Z)
+ M\omega[-{\bf X^*}\cdot {\bf r}\times {\bf Y}
  +{\bf Y^*}\cdot {\bf r} \times {\bf X}
-{\bf X^*}\cdot {\bf r}Z 
-Z^*{\bf r}\cdot {\bf X}].
\end{eqnarray*} 

From this result one deduces the equation of motion 
$$i{\partial \over \partial t}{\bf X}+({\bf \nabla} -M\omega{\bf r})\times {\bf Y}+({\bf \nabla}-M\omega {\bf r})Z=0$$
and the constraint equations for ${\bf Y}$ and $Z$
$$2M{\bf Y}+({\bf \nabla}+M\omega{\bf r})\times {\bf X}=0$$
and
$$2MZ-({\bf \nabla}+M\omega {\bf r})\cdot {\bf  X}=0.$$
This set of equations is identical in the $\omega=0$ limit to that obtained by the author  \cite{CRH} some years ago for an uncoupled Galilean vector meson.  Upon elimination of ${\bf Y}$ and $Z$ by  means of the constraint equations there results the Schr\"{o}dinger equation
$${\cal E}{\bf X}=[{{\bf p}^2\over 2M}+{1\over 2}M\omega^2{\bf r}^2-{3\over 2}\omega-\omega {\bf L}\cdot {\bf S}]{\bf X}$$
where ${\bf S}$ is the usual spin one representation of angular momentum given by 
$$(S_k)_{ij}=i\epsilon_{ijk}$$
with $\epsilon_{ijk}$ being the Levi-Civita tensor.  This result for ${\bf X}$ corresponds exactly to the Galilean spin one-half oscillator of the preceding section in that it it describes an oscillator in which the zero point energy has been eliminated and a strong spin-orbit coupling included. One also notes that the strength of that coupling corresponds exactly to a $1/S$ suppression which will in fact be seen in ${\bf IV}$ to be the general result for spin $S$. 

Similar results obtain for the spinless case although this (unsurprisingly) requires an antisymmetrical rather than a symmetrical bispinor.  Again the Lagrangian is of the form 
$${\cal L}={1\over 2}\psi^*_{ab}[G_{aa'}\Gamma_{bb'}+\Gamma_{aa'}G_{bb'}]\psi_{a'b'}$$
but in this case subject to the condition $\psi_{ab}=-\psi_{ba}$.  Such an antisymmetric spinor can be written as 
$$\psi={1\over \sqrt 2}[{\bf A}\cdot {\bf \sigma} \rho_2+B\rho_1+C{1\over 2}(1+\rho_3)]\sigma_2$$
where, analogously to the spin one case, a term proportional to ${1\over 2}(1-\rho_3)\sigma_2$ need not be included. 
The Lagrangian can, as before, be written as a trace, this time of the form
$${\cal L}=-{\rm Tr}G\psi\Gamma\psi^*.$$

Evaluation of the trace readily yields
$${\cal L}=C^*i{\partial \over \partial t}C+2M({\bf A}^*\cdot {\bf A}+B^* B)-{\bf A}^*\cdot{\bf \nabla}C+C^*{\bf \nabla}\cdot{\bf A}-M\omega[C^*{\bf  r}\cdot{\bf  A}+{\bf A}^*\cdot{\bf r}C].$$
From ${\cal L}$ one infers the vanishing of $B$, the equation of motion for $C$

$$i{\partial \over \partial t}C+({\bf \nabla}-M\omega{\bf r})\cdot{\bf A}=0,$$
and the constraint which determines ${\bf A}$ 
$$2M{\bf A}=({\bf \nabla}+M\omega {\bf r})C$$
in terms of $C$.  This again readily yields a Schr\"{o}dinger equation of the form
$${\cal E}C=[{{\bf p}^2\over 2M}+{1\over   2}M\omega^2{\bf r}^2-{3\over 2}\omega]C.$$
As in the spin one-half and spin one cases the zero point energy is absent, but, of course, there is no possibility for a spin-orbit coupling.   
\section{IV the extension to arbitrary spin}
The extension of the Galilean oscillator to the case of arbitrary spin can be accomplished by means of totally symmetrized multispinors.  To this end one employs the rank $2S$ spinor $\psi_{a_1...a_{2S}}$ where each $a_i$ ranges from 1 to 4.
In the absence of additional restrictions such an object has ${1\over 6}(2S+3)(2S+2)(2S+1)$ independent components.
The Galilean-invariant Lagrangian can be written as 
$${\cal L}={1\over 2S}\psi_{a_1...a_{2S}}^*\sum_{i=1}^{2S}\Gamma_{a_1a_1'}...\Gamma_{a_{i-1}a_{i-1}'}G_{a_ia_i'}
\Gamma_{a_{i+1}a_{i+1}'}...\Gamma_{a_{2S}a_{2S}'}\psi_{a_1'...a_{2S}'}$$
where $G$ is the usual spin one-half Galilean operator which includes the replacement of ${\bf p}$ by ${\bf p}-iM\omega{\bf r}\rho_3$.  It is to be noted that because of the occurrence of $2S-1$ matrices $\Gamma$ in the equations implied by ${\cal L}$

$$\sum_{i=1}^{2S}\Gamma_{a_1a_1'}...\Gamma_{a_{i-1}a_{i-1}'}G_{a_ia_i'}
\Gamma_{a_{i+1}a_{i+1}'}...\Gamma_{a_{2S}a_{2S}'}\psi_{a_1'...a_{2S}'}=0$$
those components of $\psi_{a_1...a_{2S}}$ for which more than one of the $a_i$ are lower indices (namely, 3 or 4) drop out of the equations.  Thus the only Galilean components are those $2S+1$ components in which all indices are 1 or 2 plus the $4S$ components in which $2S-1$ indices are 1 or 2 and one index takes the value 3 or 4.  In other words this is a $6S+1$ component theory.  Using the notation 

\begin{eqnarray*}
 \psi_{a_1...a_{2S}} & = & \phi_{a_1...a_{2S}} \;\;\;\;\;\;\;\;\;\;\;\;\;\;\; a_i=1,2 \\
\psi_{a_1...a_{2S-1}r} & = & \chi_{a_1...a_{2S-1}}^{r-2}\;\;\;\;\;\;\;a_i=1,2;\; r=3,4 
\end{eqnarray*}
the Lagrangian becomes

\begin{eqnarray*}
{\cal L} & = &
 \phi_{a_1...a_{2S}}^*i{\partial \over \partial t}\phi_{a_1...a_{2S}}+{1\over 2S}[\phi_{a_1...a_{2S}}^*\sum_{i=1}^{2S}{\bf \sigma}_{a_ir}\cdot({\bf p}+iM\omega{\bf r})\chi_{a_1...a_{i-1}a_{i+1}...a_{2S}}^r   \\
&  & + \sum_{i=1}^{2S}\chi_{a_1...a_{i-1}a_{i+1}...a_{2S}}^{r*}{\bf \sigma}_{ra_i}\cdot({\bf p}-iM\omega{\bf r})\phi_{a_1...a_{2S}}] +2M\chi_{a_1...a_{2S-1}}^{r*} \chi_{a_1...a_{2S-1}}^r.
\end{eqnarray*} 

The equation of motion is thus
$$i{\partial \over \partial t}\phi_{a_1...a_{2S}}+{1\over 2S}\sum_{i=1}^{2S}{\bf \sigma}_{a_ir}\cdot({\bf p}+iM\omega {\bf r})\chi_{a_1...a_{i-1}a_{i+1}...a_{2S}}^r=0$$
with the equation of constraint being
$$2M\chi_{a_1...a_{2S-1}}^r+ {\bf \sigma}_{ra_{2S}}\cdot ({\bf p}-iM\omega {\bf r})\phi_{a_1...a_{2S}} =0.$$
The elimination of the $\chi$ components thus yields
$$i{\partial \over \partial t}\phi_{a_1...a_{2S}}-{1\over 4MS} \sum_{i=1}^{2S}[{\bf \sigma}\cdot({\bf p}+iM\omega {\bf r}){\bf \sigma}\cdot({\bf p}-iM\omega{\bf r})]_{a_ia_{i'}}\phi_{a_1...a_{i-1}a_i'a_{i+1}...a_{2S}}=0.$$
One notes  that under an infinitesimal rotation $\delta {\bf \omega}$ the $2S+1$ component spinor $\phi_{a_1...a_{2S}}$ tranforms as a two-component spinor in each index, i.e.,  
$$\phi_{a_1...a_{2S}}'=\phi_{a_1...a_{2S}}+i\sum_{i=1}^{2S}({1\over 2}{\bf \sigma}\cdot\delta {\bf \omega})_{a_ia_{i}'}\phi_{a_1...a_{i-1}a_{i}'a_{i+1}...a_{2S}},$$
or alternatively
$$\phi'_{a_1...a_{2S}}=\phi_{a_1...a_{2S}}+i\delta {\bf \omega}\cdot
{\bf S}_{a_1...a_{2S};a_{1}'...a_{2S}'}\phi_{a_1'...a_{2S}'}$$
by definition of the total spin matrices ${\bf S}_{a_1...a_{2S};a_{1}'...a_{2S}'}$.
This allows one to write a Schr\"{o}dinger equation for $\phi_{a_1...a_{2S}}$ in  the form
$${\cal E}\phi=[{1\over 2M}{\bf p}^2+{1\over 2}M\omega^2{\bf r}^2-{3\over 2}\omega-\omega {1\over S}{\bf S}\cdot{\bf L}]\phi$$
where all matrix indices have been suppressed.  One thus finds that the generalization of the Dirac oscillator to  arbitrary Galilean spin  is simply characterized as an ordinary oscillator in which the zero point energy is absent and a spin-orbit coupling term with a $1/S$ coefficient has been included.  The eigenvalues ${\cal E}_{n\ell jS}$ for this arbitrary spin Galilean oscillator can be brought to the remarkably simple positive definite form
$${\cal E}_{n\ell jS}=\omega[2n +{(\ell+S)(\ell+S+1)-j(j+1)\over2S}]$$
where $n,\ell=0,1,2...$ and $j=\ell+S,\ell+S-1,\ell+S-2...|\ell-S|.$  One sees by inspection that the lowest state of a spin $S$ Galilean oscillator is one of zero energy and is $2S+1$ fold degenerate. 

The role played by the minimality assumption on the number of components required to describe a spin $S$ Galilean system can be displayed explicitly.  This is accomplished by using a set of components already employed in ref.[2] in the context of the $g$-factor calculation.  Specifically, one can introduce a $2S+2$-rank multispinor $\psi_{a_1...a_{2S}}^{r_1r_2}$ which is totally symmetric in all lower indices and antisymmetric in its two upper indices.  
The starting point is the Lagrangian 

\begin{eqnarray*}
{\cal L}
      &  = &  \psi_{a_1...a_{2}}^{r_1r_2*}
      \{ 
      {1\over 2}(1-\lambda)
      [\Gamma_{a_1a_{1}'}
      ...\Gamma_{a_{2S}a_{2S}'}
      (\Gamma_{r_1r_{1}'} G_{r_2r_{2}'} + G_{r_1r_{1}'} \Gamma_{r_2r_{2}'})]   \\
   &  &  \quad  + {\lambda\over 2S} \Gamma_{r_1r_{1}'} \Gamma_{r_2r_{2}'} \sum_{i=1}^{2S}\Gamma_{a_1a_{1}'}...\Gamma_{a_{i-1}a_{i-1}'} G_{a_ia_{i}'} \Gamma_{a_{i+1}a_{i+1}'}...\Gamma_{a_{2S}a_{2S}'}
    \}
      \psi_{a_1'...a_{2S}'}^{r_1'r_2'} 
  \end{eqnarray*}    
where $\lambda$ is an arbitrary real parameter.
Suppressing for the moment all lower indices on $\psi$, one can write as in the spin zero case
$$\psi^{r_1r_2}={1\over \sqrt 2}\{[{1\over 2}(1+\rho_3)\phi+i\psi^k \sigma_k\rho_2+\psi'\rho_1]\sigma_2\}_{r_1r_2}$$
where $\psi$, $\psi_k$, and $\psi'$ are all multispinors of rank $2S$.   As already seen in the spinless case one can omit a term proportional to $(1-\rho_3)\sigma_2$ and can also drop the $\psi'$ term as it makes no contribution. 

The part of ${\cal L}$ proportional to $1-\lambda$ can be written as 
$$-(1-\lambda){\rm Tr}G\psi\Gamma\psi^*$$
with only the components for which $a_i=1,2$ in the $2S$ symmetrized indices of $\phi$ and $\psi^k$ being relevant because of the $2S$ factors of $\Gamma_{a_ia_i'}$.  There are therefore apparently $4(2S+1)$ components which survive in the above trace.  Its evaluation yields
$$\phi^*i{\partial \over \partial t}\phi-{\bf \psi}^*\cdot({\bf p}-iM\omega{\bf r})\phi-\phi^*({\bf p} +iM\omega{\bf r})\cdot\psi+2M{\bf \psi}^*\cdot{\bf \psi}.$$

The part of ${\cal L}$ proportional to $\lambda$ is similarly calculated but in this case only that part of $\psi^{r_1r_2}$
proportional to $\phi$ contributes.  One now calls $\phi$ only that part in which all $a_i=1,2$, and those components of $\phi$ for which one component is 3 or 4 are denoted by $\chi_{a_1...a_{2S-1}}^r$ where $r\equiv a_{2S}-2$.  One finds that the total Lagrangian becomes 

\begin{eqnarray*}
{\cal L} & = &
 \phi_{a_1...a_{2S}}^*i{\partial \over \partial t}\phi_{a_1...a_{2S}}+(1-\lambda)[\phi_{a_1...a_{2S}}^*({\bf p}+iM\omega{\bf r})_k\psi_{a_1...a_{2S}}^{k}
  +\psi_{a_1...a_{2S}}^{k*}({\bf p}-iM\omega{\bf r})_k\phi_{a_1...a_{2S}}\\
& &+2M\psi_{a_1...a_{2S}}^{k*}\psi_{a_1...a_{2S}}^k] 
+ {\lambda\over 2S}[\phi_{a_1...a_{2S}}^*\sum_{i=1}^{2S}{\bf \sigma}_{a_ir}\cdot({\bf p}+iM\omega{\bf r})\chi_{a_1...a_{i-1}a_{i+1}...a_{2S}}^r   \\
&  & + \sum_{i=1}^{2S}\chi_{a_1...a_{i-1}a_{i+1}...a_{2S}}^{r*}{\bf \sigma}_{ra_i}\cdot({\bf p}-iM\omega{\bf r})\phi_{a_1...a_{2S}}] +2M\chi_{a_1...a_{2S-1}}^{r*} \chi_{a_1...a_{2S-1}}^r
\end{eqnarray*} 
where all spinor summations are now over two-valued indices.  

The equations implied by ${\cal L}$ are thus seen to be 
$$i{\partial \over \partial t}\phi_{a_1...a_{2S}}+(1-\lambda)({\bf p}+iM\omega{\bf r})\cdot{\bf \psi}_{a_1...a_{2S}}+{\lambda\over 2S}\sum_{i=1}^{2S}{\bf \sigma}_{a_ir}\cdot({\bf p}+iM\omega {\bf r})\chi_{a_a...a_{i=1}a_{i+1}...a_{2S}}^r=0$$

$$2M\psi_{a_1...a_{2S}}^k+(p_k-iM\omega r_k)\phi_{a_1...a_{2S}}=0$$
and 
$$2M\chi_{a_1...a_{2S-1}}^r+{\bf \sigma}_{ra_{2S}}\cdot({\bf p}-iM\omega {\bf r})\phi_{a_1...a_{2S}}=0.$$
These yield the modified Schr\"{o}dinger equation 
$$E\phi=[{1\over 2M}{\bf p}^2+{1\over 2}M\omega^2{\bf r}^2-{3\over 2}\omega-\omega{\lambda\over S}{\bf S}\cdot{\bf L}]\phi=0$$
thereby demonstrating that this $12S+4$ component theory is sufficiently general to yield an arbitrary coefficient for the spin-orbit coupling term.  It is worth noting that this allows even the spin one-half Galilean oscillator to be extended to a ten component theory which could have an arbitrary coefficient for this term.  

\section{V conclusion}
It has been demonstrated in this paper that the Dirac oscillator can be extended beyond the spin one-half case to that of arbitrary spin within the framework of Galilean relativity and that this generalization leads to an oscillator system which does not have a  zero point energy but does include a spin-orbit coupling term.  Special relativity may require the existence of new physical effects so long as they have the property of vanishing in the Galilean limit.  In this context it should be noted that the attempt by Moshinsky and del Sol Mesa \cite{Sol} 
 to find an arbitrary spin version of the Dirac oscillator within the framework of special relativity cannot be considered to have succeeded since it does not in fact have a Galilean limit.  The Galilean results presented here could provide the basis for an extension into the realm of special relativity  which might well  contain additional interaction terms.



\end{document}